\documentclass[aip,reprint,jcp]{revtex4-1}

\usepackage{graphicx}
\usepackage{color}
\usepackage{dsfont}

\DeclareMathAlphabet{\mathpzc}{OT1}{pzc}{m}{it}

\newcommand{\RNum}[1]{\uppercase\expandafter{\romannumeral #1\relax}}

\begin{document}


\title{Dynamical States in Driven Colloidal Liquid Crystals} %Title of paper



\author{Ellen Fischermeier}
 \email[]{ellen.fischermeier@fau.de}
%\thanks{}
\author{Matthieu Marechal}
\author{Klaus Mecke}
\affiliation{Institut f\"{u}r Theoretische Physik, Friedrich-Alexander Universit\"{a}t Erlangen-N\"{u}rnberg,  D-91058 Erlangen, Germany}


\date{\today}

\begin{abstract}
We study a model colloidal liquid crystal consisting of hard spherocylinders under the influence of an external aligning potential by Langevin dynamics simulation. 
The external field that rotates in a plane acts on the orientation of the individual particles and induces a variety of collective nonequilibrium states. 
We characterize these states by the time-resolved orientational distribution of the particles and explain their origin using the single particle behavior. 
By varying the external driving frequency and the packing fraction of the spherocylinders we construct the dynamical state diagram.
\end{abstract}


\maketitle %\maketitle must follow title, authors, abstract and \pacs


% section Introduction
\section{Introduction\label{sec:introduction}}
The manipulation of liquid crystals by external fields is a key mechanism in electro-optical applications~\cite{Hoogboom_LCD:2006}. Therefore, it is not surprising that this has inspired a lot of research, 
both experimentally and theoretically. 
In this work we investigate the dynamical states that evolve in a colloidal liquid crystal under the influence of a rotating aligning potential. The advantage of colloidal over molecular systems lies in the fact 
that they can be directly studied in real space giving access to the dynamics of individual particles, e.g. Ref.~\onlinecite{Lettinga_Hydrodynamic:2010,Kuijk_Phase:2012}.\\
%Liquid crystals can be broadly classifies into two types: materials consisting of either anisotropic molecules or anisotropic colloidal particles. The advantage 
%of the latter is the fact that they can be directly studied in real space, giving access to the dynamics of individual particles, e.g.~\cite{Lettinga_Hydrodynamic:2010,Kuijk_Phase:2012}.\\
There have been a number of studies on the field-induced behavior of %individual 
isolated anisotropic colloidal particles, such as 
optically torqued nanorods~\cite{Shelton_Nonlinear:2005}, the electrorotation of nanowires~\cite{Edwards_Synchronous:2006},
and magnetic rods in precessing magnetic fields~\cite{Dhar_Orientations:2007,Tierno_Overdamped:2009,Coq_Three-dimensional:2010}.
These studies agree that for low frequencies 
the particle dynamics show a linear regime in which the orientations of the particles simply follow the external field. 
For higher frequencies, however, the dynamics become nonlinear showing for example a periodic
switching in rotation direction~\cite{Shelton_Nonlinear:2005,Tierno_Overdamped:2009}.
%
%For colloidal many particle systems, studies on spherical particles imposed an anisotropy by a permanent dipole moment of the particles~\cite{Murashov_Structure:2000,Jager_Pattern:2011}.
In colloidal many-particle systems, the collective behavior depends critically on the shape of the particles and the type of time-dependent external field. 
For example spheres with a permanent dipole assemble into layers under a rotating magnetic field~\cite{Murashov_Structure:2000,Jager_Pattern:2011}.
%Here the interplay of interaction with a rotating magnetic field and dipole-dipole interactions leads to layer formation in the field plane. %\\
In systems of platelets between two flat parallel walls, rotating magnetic fields have been experimentally shown to induce a bend-splay Frederiks transition with transient spatially 
periodic patterns~\cite{Beek_Influence:2008}. 
For rod-like particles, a linearly oscillating electric field will lead to a number of complex states and phases consisting of isotropic, nematic and chiral nematic domains~\cite{Kang_Double-layer:2008}.\\
We study the arguably most simple model system for a liquid crystal~\cite{Bolhuis_Tracing:1997} under the influence of the external field proposed by H\"artel et al.~\cite{Haertel_Towing:2010}. Here the colloidal particles are 
represented by hard spherocylinders of aspect ratio $L/D=5$ (see Fig.~\ref{fig:sketch}) which are subjected to an aligning field that rotates in a plane. In their dynamical
fundamental measure density-functional theory study H\"artel et al. observed a variety of dynamical states, depending on the packing fraction and driving frequency of the field. 
%The aim of our Langevin dynamics 
%simulation is to reevaluate these states by analyzing the behavior of individual particles which is not possible in density-functional theory. 
In light of recently discovered shortcomings~\cite{Krueger_shear_DDFT} of dynamical density functional theory (DDFT), it is worthwhile
to revisit this system with computer simulations. Any observed deviations from DDFT would then also lend
further support to recently proposed improvements of the theory~\cite{Schmidt_power_functional,Krueger_shear_II}.
In addition, our Langevin dynamics simulations allow us to access the single-particle behavior, which is not possible
in (dynamical) density functional theory. We are thus able to give a conclusive analysis of the origin
and the underlying mechanisms of the different states.\\
The paper is structured as follows: In Sec.~\ref{sec:numerical} we give the details of the simulation method and the system setup. The external potential is discussed in Sec.~\ref{sec:external}, the results
we obtain are presented in Sec.~\ref{sec:results} and we summarize our conclusions in Sec.~\ref{sec:conc}. 

% section on numerical method
\section{Simulation method\label{sec:numerical}}
\begin{figure}[h!tbp]
 \centering
  \includegraphics{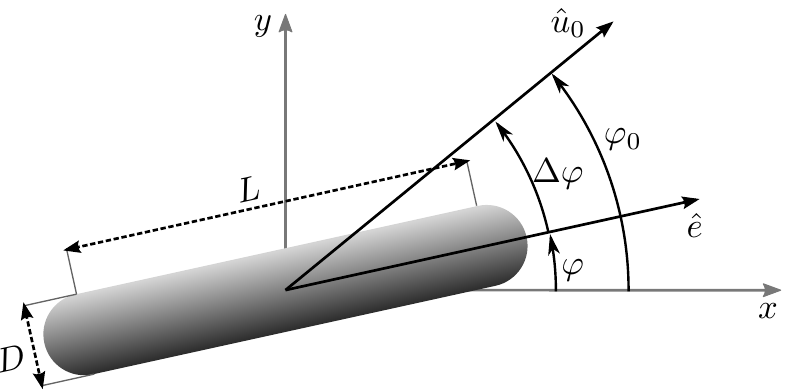}
  \caption{Sketch of a spherocylinder with aspect ratio $L/D=5$ in the equatorial plane. Its orientation is given by the normalized orientation vector $\hat{e}$; 
          $\hat{u}_0$ indicates the alignment direction in which the external potential is minimal.\label{fig:sketch}}
\end{figure}

In this work we model the colloidal system via a Langevin dynamics simulation where the solvent is not explicitly included. The equations of motion for translation and rotation can be written 
as~\cite{Doi_Theory:1998}:

\begin{eqnarray}
\label{momentum}
 \frac{d\vec{p}}{dt}&=&-\mathbf{\Xi} \vec{v} +\vec{F}_S +\vec{F}_R\\
\label{angular_momentum}
 \frac{d\vec{L}}{dt}&=&-\gamma_r \vec{\omega} +\vec{T}_S +\vec{T}_R
\end{eqnarray}
where $\vec{p}=m\vec{v}$ and $\vec{L}=\mathbf{I}\vec{\omega}$ are the momentum and angular momentum of a particle with mass $m$ and inertia tensor $\mathbf{I}$ with $\vec{v}$ and $\vec{\omega}$ 
its translational and angular velocity.\\
The first term on the right hand side of Eq.~{(\ref{momentum})} and (\ref{angular_momentum}) accounts for viscous dissipation. The translational friction tensor $\mathbf{\Xi}$ depends on the 
translational friction coefficients 
$\gamma_{\|}$ and $\gamma_{\bot}$ for motion parallel and perpendicular to the symmetry axis
$\hat{e}$ of particle $i$: %~\cite{Tao_Isotropic:2006}, $\delta_{\alpha\beta}$ being the Kronecker delta
\begin{equation}
 \mathbf{\Xi}_i=\gamma_{\|}\hat{e}_i\otimes\hat{e}_i+\gamma_{\bot}(\mathds{1}-\hat{e}_i\otimes\hat{e}_i)%\Xi_{\alpha\beta}=\gamma_{\|}e_{\alpha}e_{\beta}+\gamma_{\bot}\left(\delta_{\alpha\beta}-e_{\alpha}e_{\beta}\right)
\end{equation}
The translational friction coefficients and the rotational friction coefficient $\gamma_r$ are linked to translational and rotational diffusion via the  Einstein-Smoluchowski relation $\gamma_i=k_B T/ D_i$ 
with $k_B$ as Boltzmann's constant, $T$ the Temperature and $D_i$ the respective diffusion constant. They only depend on size and aspect ratio of the particle and the viscosity of the fluid~\cite{Tirado_Comparison:1984}.\\
The subscripts $R$ and $S$ in Eq.~{(\ref{momentum})} and (\ref{angular_momentum}) indicate the systematic and random contribution to the force $\vec{F}$ and torque $\vec{T}$, respectively.
The random contributions are related to the friction according to the fluctuation dissipation theorem for particles $i$ and $j$:

\begin{eqnarray}
 \langle\vec{F}_{R,i}(t)\rangle&=&\langle\vec{T}_{R,i}(t)\rangle=\vec{0}\\
 \langle\vec{F}_{R,i}(t)\vec{F}_{R,j}(t')\rangle&=&2 k_B T \mathbf{\Xi}_i\,\delta_{ij} \delta(t-t')\\
 \langle\vec{T}_{R,i}(t)\vec{T}_{R,j}(t')\rangle&=&2 k_B T \gamma_r(\mathds{1}-\hat{e}_i\otimes\hat{e}_i)\,\delta_{ij}\delta(t-t')
\end{eqnarray}
with $\delta(t-t')$ the Dirac delta distribution which in the case of discrete time steps of size $\Delta t$ is replaced by $\delta_{tt'}/\Delta t$. 
For reasons of symmetry we keep $\vec{\omega}$ perpendicular to the symmetry axis of the spherocylinder given by the normalized 
orientation vector $\hat{e}$ (see Fig.~\ref{fig:sketch}). We therefore only apply random torques $\vec{T}_R$ perpendicular to this axis.\\
The systematic contributions originate from both the torque exerted by the external potential (see Sec.~\ref{sec:external}) and the particle particle interactions 
which in our model only arise from excluded volume, i.e., 
electrostatic interactions between colloids are neglected as they are assumed to be sufficiently screened by the ions in the solvent.
We use a rigid body dynamics framework~\cite{Iglberger_Rigid:2009,Iglberger_Software:2010} to obtain the motion of the spherocylinders according to Eq.~{(\ref{momentum})} and (\ref{angular_momentum}) 
and to resolve collisions between particles. 
To get the correct friction coefficients we previously performed drag tests on the spherocylinders in a Lattice Boltzmann simulation and found them to agree quite well with the values given by 
Tirado et al.~\cite{Tirado_Comparison:1984} for cylinders with an aspect ratio of $5.7$ at the same diameter. This can be understood by the fact that, while the cylindrical part of the spherocylinders has 
an aspect ratio of five, their total aspect ratio including the end caps is six.\\
%\todo{$\varphi$, $\Delta\varphi$, $\omega_0 t$?}\\
The hard spherocylinder systems we evaluate in this work consist of between $2304$ and $9126$ particles %with an aspect ratio of $L/D=5$ 
at different number densities $\rho$. 
Throughout this work number densities are given as dimensionless densities $\rho^*=\rho/\rho_{cp}$ normalized to the density in the close packing limit 
$\rho_{cp}=(2 D^{-3})/(\sqrt{2}+\frac L D \sqrt{3})$. 
The particle mass and inertia were chosen such that the timescales $\tau_{{ac}_t}=m/\gamma_t$ and $\tau_{ac_r}=I_{\bot}/\gamma_r$ on which the translational and angular velocity autocorrelations decay are 
much smaller than the corresponding Brownian timescales $\tau_{b}$. Here $\gamma_t$ signifies the translational friction coefficient for motion in a random direction given by $[2/3\gamma_{\bot}+1/3\gamma_{\|}]^{-1}$ according to Ref.~\onlinecite{Tirado_Comparison:1984} 
and $I_{\bot}$ is the moment of inertia perpendicular to $\hat{e}$. If we define the Brownian timescales as $\tau_{{b}_t}=L^2\gamma_t/6k_B T$ and $\tau_{{b}_r}=(\pi/2)^2\gamma_r/4k_B T$ we obtain ratios of 
$\tau_{{ac}_t}/\tau_{{b}_t}=1.8\times 10^{-4}$ and $\tau_{{ac}_r}/\tau_{{b}_r}=1.0\times 10^{-4}$. This difference in timescales is chosen to reduce the impact of inertia on our results.
Furthermore, we use periodic boundary conditions in our simulations to minimize finite size effects.\\

% theoretical Background
\section{External potential\label{sec:external}}
We investigate the dynamical states that evolve in the model liquid crystal under the influence of the external potential:
\begin{equation}
V_{ext}(t,\hat{e})=-V_0\left<\hat{u}_0(t),\hat{e}\right>^2=-V_0 \cos^2(\omega_0 t-\varphi)\sin^2(\vartheta)
\end{equation}
This potential acts on the orientation $\hat{e}$ of each individual particle, where $\varphi$ and $\vartheta$ are its azimuthal and polar 
angle in spherical coordinates.
It favors an alignment of the symmetry axis of the spherocylinder $\hat{e}$ to a direction $\hat{u}_0(t)=\pm (\cos\omega_0 t, \sin \omega_0 t,0)=\pm (\cos\varphi_0, \sin \varphi_0,0)$
that rotates in the $xy$-plane with a given frequency $\omega_0$. The $\pm$ originates from the symmetry
of the particles that makes $\hat{e}$ equivalent to $-\hat{e}$ (see Fig.~\ref{fig:sketch}). In experiments, this could be realized by a high frequency electric field oriented along $\hat{u}_0(t)$
that induces dipole moments in the colloid particles~\cite{Edwards_Synchronous:2006}. However, the resulting dipole-dipole interactions are absent in our simulation.\\
Throughout this work we keep the strength of the potential fixed at $V_0=5\,k_B T$ and vary the rotation frequency $\omega_0$. %the density $\rho^*$ and 
\subsection{Single particle dynamics in the absence of noise}
\label{subsec:single}
Assuming an isolated spherocylinder aligned to the equatorial plane, the torque exerted on it depends on the phase $\Delta\varphi =\omega_0 t-\varphi$ between the most favorable alignment direction 
$\hat{u}_0$ and the orientation of the particle $\hat{e}$ (see Fig.~\ref{fig:sketch}) as:
\begin{equation}
\label{torque}
 \vec{T}_{ext}=V_0\sin(2\Delta\varphi)\hat{e}_z
\end{equation}
The torque balance in equilibrium between this external torque $\vec{T}_{ext}$ and the rotational friction torque $-\gamma_r \vec{\omega}$ gives rise to a critical frequency of the external potential 
$\omega_0^*=V_0/\gamma_r$. Up to this frequency the particle 
orientation $\hat{e}$ will rotate with the same frequency as the external potential and the phase $\Delta\varphi$ between them is constant at:
\begin{equation}
 \Delta\varphi=\frac 1 2 \arcsin(\frac{\gamma_r}{V_0}\omega_0)
\label{eq:deltaphi}
\end{equation}
With increasing $\omega_0$ this phase shift increases until for 
$\omega_0^*$ a phase shift of $\pi/4$ is reached at which the external torque is maximal. For frequencies above $\omega_0^*$ there exists no solution with a constant phase shift as the corresponding 
friction would exceed the maximum torque the external potential can exert. Instead $\Delta\varphi$ will increase continuously although it can of course always be mapped into the interval of $[-\pi/2,\pi/2[$ 
due to the symmetry of the particles. For $0<\Delta\varphi<\pi/2$ the particles are subjected to a torque that turns them in the same direction as the field is turning while for $-\pi/2<\Delta\varphi<0$ the 
torque acts in the opposite direction.
For a detailed discussion of the nonlinear differential equation of motion see Ref.~\onlinecite{Shelton_Nonlinear:2005,Edwards_Synchronous:2006}.
Throughout this paper we will give all frequencies as multiples of the critical frequency $\omega_0^*$.
\subsection{Limiting cases of the many-particle system}
\begin{figure}[h!tbp]
 \centering
  \includegraphics{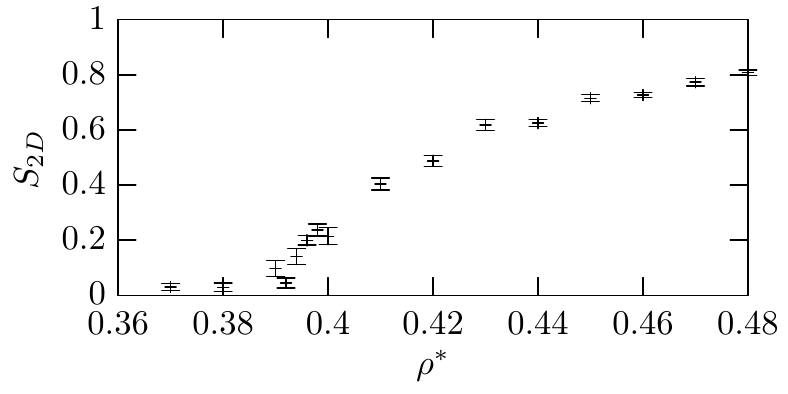}
  \caption{Two-dimensional order parameter $S_{2D}$ over dimensionless density $\rho^*$\label{fig:phasetransition}}
\end{figure}

%From density functional theory studies~\cite{Haertel_Towing:2010,Wensinki_Nematic:2005} we expect two find two distinct regimes in density, above and bellow $\rho_c^*=0.3717$\todo{check numerically?}, the density at which the system system undergoes a paranematic-nematic 
%transition for $\omega_0\rightarrow\infty$. 
In the limiting case of infinite driving frequency $\omega_0\rightarrow\infty$ the spherocylinders feel an effective time-averaged potential $-V_0\sin^2(\vartheta)/2$. 
From density-functional theory studies~\cite{Haertel_Towing:2010,Wensinki_Nematic:2005} we expect to find two distinct regimes in density separated by a paranematic to nematic phase transition. %two distinct regimes in density, above and bellow $\rho_c^*=0.3717$
We characterize the system by the two-dimensional order parameter $S_{2D}$ defined as the larger eigenvalue of the two-dimensional ordering tensor
\begin{equation}
 Q_{\alpha\beta}=\frac 1 N \left(\sum_i \frac {2}{\hat{e}_{ix}^2+\hat{e}_{iy}^2} \hat{e}_{i\alpha} \hat{e}_{i\beta}\right)- \delta_{\alpha\beta} \,\, , \,\,\, \,\,\,\alpha,\beta=x,y
\label{eq:orderparam}
\end{equation}
Our simulations confirm the phase transition reported in Ref.~\onlinecite{Haertel_Towing:2010} for this system. We estimate the transition to occur at a dimensionless density of about $\rho_c^*=0.39$ 
(Fig.~\ref{fig:phasetransition}). This value is slightly larger than 
the one given in Ref.~\onlinecite{Haertel_Towing:2010}.
For densities lower than $\rho_c^*$ the system exhibits a
paranematic phase where the orientational distribution of the particles will be isotropic in $\varphi$ but peaked strongly around $\vartheta=\pi/2$. For densities above $\rho_c^*$ an additional ordering 
arises in $\varphi$ leading to a nematic phase with the director lying in the equatorial plane.\\
In the opposite limit of vanishing driving frequency $\omega_0=0$ the external potential induces a nematic state with the director aligned to the $x$-axis for all densities.
\\%\todo{$\varphi$, $\Delta\varphi$, $\omega_0 t$?}\\
%

% section Results
\section{Results}
\label{sec:results}
%
%dimensionless density of $\rho^*=0.5$ 
%where the system is expected to be in the nematic phase~\cite{Frenkel_Structure:1988,Frenkel_Thermodynamic:1988,Veerman_Phase:1990,Bolhuis_Tracing:1997,McGrother_re‐examination:1996} 
%with $\rho^*=\rho/\rho_{cp}$, $\rho_{cp}$ being the density in the close packing limit $\rho_{cp}=(2 D^{-3})/(\sqrt{2}+\frac L D \sqrt{3})$.\\
%
To sample the space of dynamical states that evolve in the system under the influence of the external potential described in Sec.~\ref{sec:external}, we either varied the driving frequency $\omega_0$ at 
fixed density $\rho^*$ or changed $\rho^*$ at constant $\omega_0$. 
In the first case the systems were first equilibrated in the absence of an external potential. After equilibrium was reached, the external potential was switched on and the system was again allowed to 
equilibrate until a dynamical steady state evolved. In the second case a system with a fully developed dynamical steady state at $\rho^*=0.48$ and $\omega_0=0.8\,\omega_0^*$ was used as initial 
configuration. The density was then successively lowered by rescaling the particle positions. For each new packing fraction the system again was allowed to relax to the corresponding dynamical state.\\
In the following 
%section
we will describe the dynamical states we found in detail. We focus on the angular distribution of the particle orientation vectors $\hat{e}$
and on the height $I_p$ and orientation $(\varphi_p,\vartheta_p)$ of the peak with the highest intensity of this distribution. For a nematic phase this height would be closely linked to the order parameter, 
while the orientation would 
correspond to the nematic director.\\
As the time independent part of the external potential $V_{ext}(\vartheta)=-V_0\sin^2(\vartheta)$ ensures that the orientation of the peak lies in the equatorial plane, i.e., $\vartheta_p=\pi/2$, 
we can reduce this 
evaluation to tracking the azimuthal angle $\varphi_p$ of the peak orientation. Furthermore, we average the angular distribution over the two representatives $\hat{e}$ and $-\hat{e}$ 
of the particle orientation. Thus the distribution is periodic in $\varphi$ with a periodicity of $\pi$.\\
Movies of the time evolution of the azimuthal angular distribution for the different dynamical states are also available~\cite{supplemental}.
\subsection{High density regime ($\rho^*>\rho_c^*$)}
\begin{figure}[ht]
  \includegraphics[]{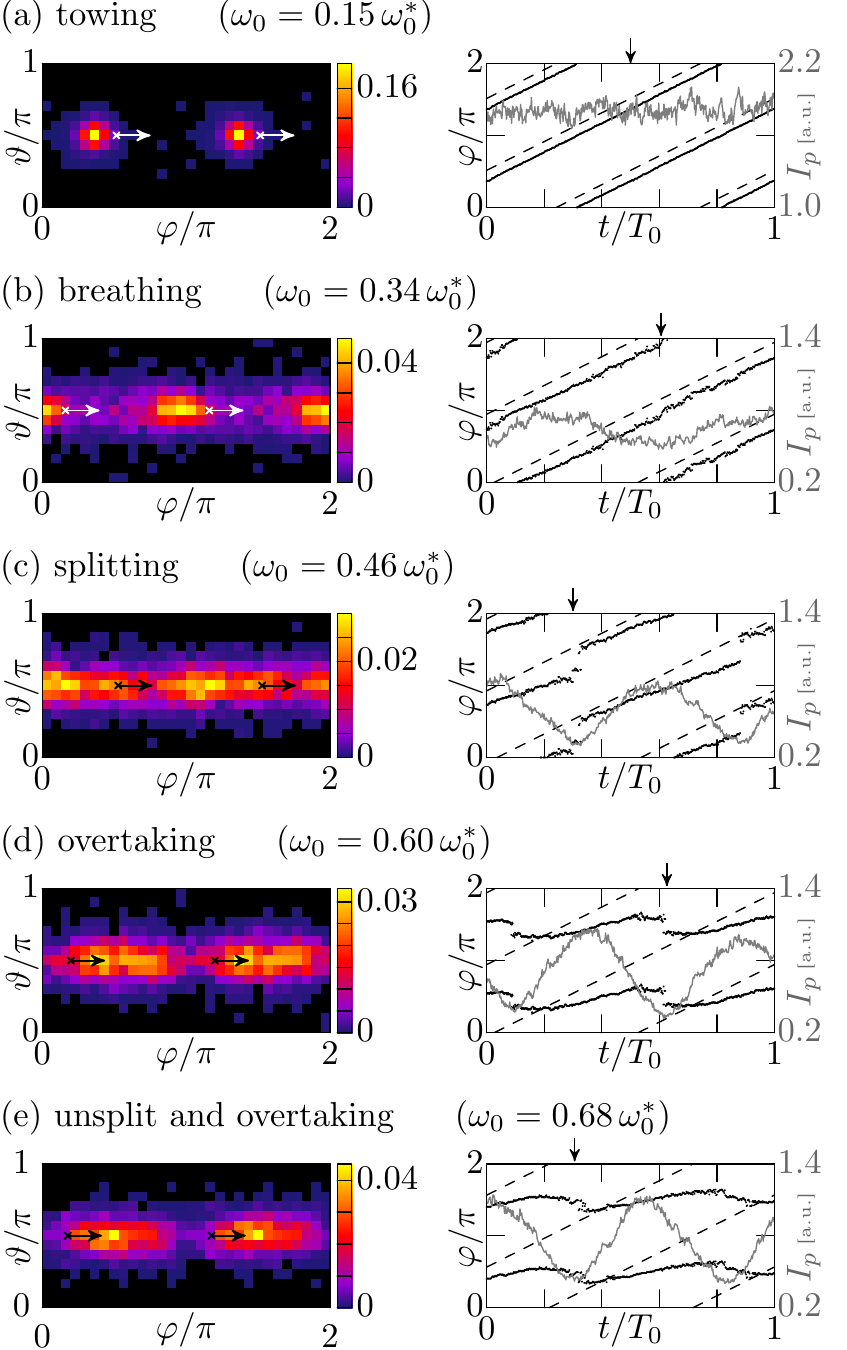}
\caption{Dynamical states evolving at $\rho^*=0.48$. Left: Snapshot of the angular distribution of the particle orientations taken at the time indicated by the arrow in the right figure. 
The color scale is normalized to the total number of particles.
 The cross marks the orientation of the potential minimum $\hat{u}_0(t)$, the arrow the turning direction. 
 Right: Azimuthal angle $\varphi_0$ of the potential minimum (dashed line) and of the maximum peak in the angular distribution $\varphi_p$ (black dots) as well as
the height $I_p$ of this peak (gray line) over one period of the external potential.\label{fig:statesNem}}
\end{figure}

For systems at high densities we discover a broad variety of dynamical states as depicted in Fig.~\ref{fig:statesNem}.%
\subsubsection*{towing} 
At low frequencies (Fig.~\ref{fig:statesNem}a), the individual particles are able to rotate
at the same frequency as the external field. For the angular distribution this results in a nematic peak with constant intensity $I_p$. However, the phase difference 
$\Delta\varphi_p=\varphi_0-\varphi_p$ is 
larger in the many-particle system than the phase shift expected for an isolated particle according to Eq.~{(\ref{eq:deltaphi})}, 
since particle-particle interactions provide an additional source of friction. As the peak is 
dragged behind $\hat{u}_0$ at a constant phase shift $\Delta\varphi_p$, this state is called towing~\cite{Haertel_Towing:2010}.
\subsubsection*{breathing}
Upon increasing the driving frequency particle-particle interactions make it impossible for some particles to follow the external potential as they hinder each other in their turning.
%Before they can rearrange by diffusion, some of them acquire a phase shift $\Delta\varphi$ too large for the friction to be balanced by the external torque. %such that they are able to pass each other,
As they get left behind, the intensity $I_p$ of the peak decreases and $\Delta\varphi_p$ increases slightly (see Fig.~\ref{fig:statesNem}b). If $\Delta\varphi$ of the stray particles becomes 
larger than $\pi/2$ 
they are attracted to the subsequent potential minimum and change their turning direction. When they are overtaken by the potential minimum at $\varphi_0-\pi$ they again reverse their rotation direction
 and are collected into the peak at $\varphi_p-\pi$. 
As this peak is equivalent to the one at $\varphi_p$, as mentioned above, this leads to a re-increase in $I_p$ and a decrease in 
$\Delta\varphi_p$. Due to this periodic change in $I_p$ and $\Delta\varphi_p$ this state is called breathing. The breathing frequency itself is larger than $\omega_0$ and increases slightly with increasing 
driving frequency.
\subsubsection*{splitting}
With further increase in the external rotation frequency (Fig.~\ref{fig:statesNem}c) the spherocylinders have less and less time to rearrange amongst each other. Thus their orientations lag 
further and further behind $\hat{u}_0$ and with them the peak orientation $\varphi_p$. If $\omega_0$ surpasses a certain value, $\Delta\varphi_p$ becomes larger than $\pi/4$. Thus the torque on
the majority of the particles starts to decrease (see Eq.~\ref{torque}) which leads to an even faster increase in $\Delta\varphi_p$. At the same time 
those particles that broke away from the main peak first are overtaken by the potential minimum at $\varphi_0-\pi$ like in the breathing case. The torque on them increases until at some phase shift %subsequent minimum of the external potential
$\Delta\varphi\in\left]\pi,\pi+\pi/4\right[$, which is equivalent to a phase shift of $\Delta\varphi\in\left]0,\pi/4\right[$, they experience a stronger torque than those particles constituting the 
original peak at its current $\Delta\varphi_p\in\left]\pi/4,\pi/2\right[$. 
Thus they become no longer integrated into the main peak, % at $\varphi_p-\pi$
but instead give rise to the formation of a second peak while the original peak continues to drop in intensity. As we track the 
orientation of the peak with the highest intensity, we see a discontinuity in $\varphi_p$ when the newly formed peak surpasses the old one in intensity. This state is called splitting due to the fact 
that there is no longer only one peak on each hemisphere but two.
\subsubsection*{overtaking}
For even higher frequencies the modulus of the jump in azimuthal angle from the decaying peak to the one originating from particles left behind %by the peak at $\varphi_p+\pi$ 
becomes smaller than $\pi/2$ 
(Fig.~\ref{fig:statesNem}d). 
The reason for this is that with increasing $\omega_0$ the phase shift $\Delta\varphi_p$ increases faster while 
at the same time particles that lag even further behind no longer have time to reorient very far before they are dragged again by the potential minimum at $\varphi_0-\pi$.
Following Ref.~\onlinecite{Haertel_Towing:2010} the branches of $\varphi_p$ in both 
Fig.~\ref{fig:statesNem}c) and Fig.~\ref{fig:statesNem}d) can each be combined to a unique
function $\varphi_p(t)$ such that the modulus of the jumping angle is always smaller than $\pi/2$. This would make the splitting state correspond to a peak jumping forwards, 
while in the case of Fig.~\ref{fig:statesNem}d) the peak in the angular distribution is overtaken by 
the potential minimum once during each breathing cycle of $I_p$ hence the name overtaking.
\subsubsection*{unsplit and overtaking}
Finally, above a certain frequency $\omega_0$, we again observe only one peak (Fig.~\ref{fig:statesNem}e). Here the orientation of the majority of the particles falls rather quickly behind $\hat{u}_0$. 
As soon as $\Delta\varphi_p$ becomes larger than $\pi/4$ the aligning torque on the main part of the spherocylinders decreases leading to a broadening of the peak accompanied by a drop in intensity $I_p$. 
As in the cases above, particles that gain a $\Delta\varphi$ larger than $\pi/2$ are attracted to the minimum at $\varphi_0-\pi$ and change their turning directions. 
But before they become fully separated from the main peak, $\Delta\varphi_p$ itself 
becomes larger than $\pi/2$. Thus the whole peak starts to turn in the opposite direction as the external potential. At the same time those particles with the largest $\Delta\varphi$, i.e., those 
that are overtaken by the potential minimum at $\varphi_0-\pi$, again change their turning direction and are pushed back towards the peak leading to a re-increase in $I_p$. When the peak itself is overtaken 
by the potential minimum at $\varphi_0-\pi$ it too reverses its turning direction. While the peak is dragged by the external potential it collects those particles with $\Delta\varphi<\Delta\varphi_p$ gaining further in intensity 
$I_p$ until above $\Delta\varphi_p=\pi+\pi/4$ the process repeats itself. As there is now only one peak that is periodically overtaken by the preferred direction of alignment this state is known as 
unsplit and overtaking. 
%
%fig_slim
%
\subsection{Low density regime ($\rho^*<\rho_c^*$)}
\begin{figure}[ht]%\vspace{-3.5cm}
  \includegraphics[]{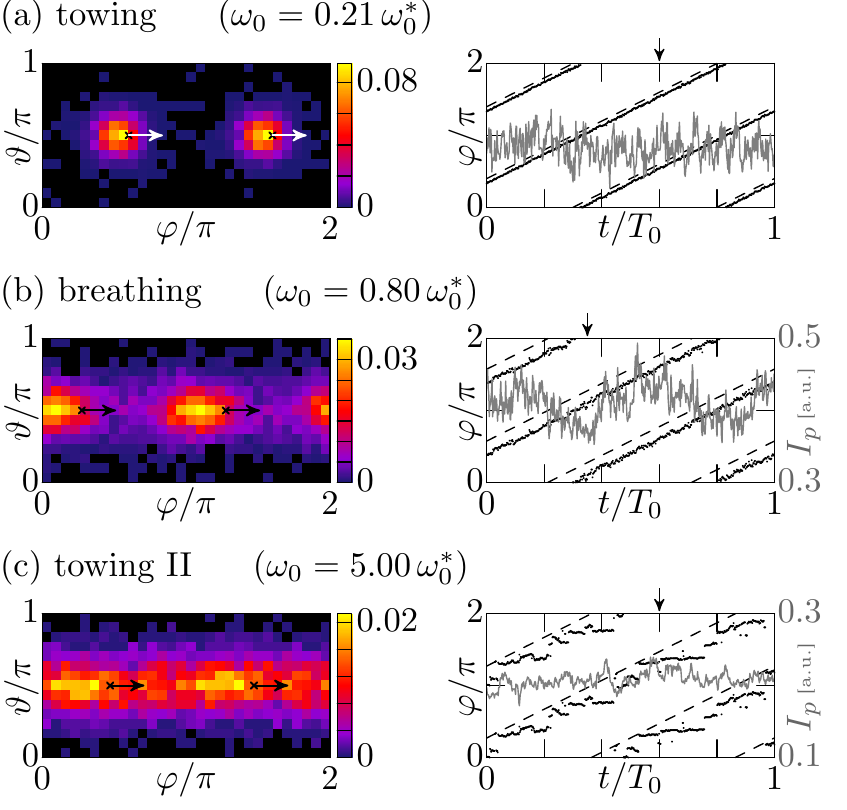}
\caption{Dynamical states at $\rho^*=0.25$. Symbols as in Fig.~\ref{fig:statesNem}\label{fig:statesParanem}}
\end{figure}

In the low density regime we observe three dynamical states that are depicted in Fig.~\ref{fig:statesParanem}.
\subsubsection*{towing} 
At low frequencies (Fig.~\ref{fig:statesParanem}a), we find again the behavior described above as towing. Here the difference between $\Delta\varphi_p$ and the phase shift in the isolated 
particle case is not as prominent as in the high particle density regime as particle-particle interactions are less dominant.
\subsubsection*{breathing}
For intermediate driving frequencies (Fig.~\ref{fig:statesParanem}b) the system develops a dynamical breathing state that corresponds qualitatively to the one observed for higher packing fractions. 
However, as the inter-particle interactions are less pronounced than at high particle densities the transition between towing and breathing occurs at higher driving frequencies than in the denser 
packed system. Additionally this results in both a lower peak intensity and a lower breathing amplitude than in Fig.~\ref{fig:statesNem}b).
\subsubsection*{towing~\RNum{2}}
At frequencies beyond the critical frequency the individual particles can not follow $\hat{u}_0$. Their orientations are rather isotropically distributed in $\varphi$ as the system is driven towards the 
paranematic state. The peak visible in 
Fig.~\ref{fig:statesParanem}c) arises from those particles that have a $\Delta\varphi$ of close to $\pi/4$ towards the current orientation of $\hat{u}_0$ and thus feel a stronger torque than the other
particles. They start to rotate with the external potential until friction becomes dominant, contributing for that time to a peak in the angular distribution that is following $\hat{u}_0$. This is a rather 
subtle effect as can be seen by the strong noise in $\Delta\varphi_p$ and the intensity scale of Fig.~\ref{fig:statesParanem}c) in comparison to Fig.~\ref{fig:statesParanem}a). 
%Therefore Fig.~\ref{fig:statesParanem}b) had to be comprised of the time average over six independent simulations. 
We expect the peak intensity to continuously decrease further with increasing $\omega_0$ until it vanishes for $\omega_0\rightarrow\infty$ in 
the paranematic state. To discriminate this purely collective soliton-like phenomenon from the synchronized turning of individual particles we call it towing~\RNum{2}.\\
\subsection{Nonequilibrium state diagram} %and comparison to dynamical fundamental measure density-functional theory
\begin{figure}[h]
  \centering
  \includegraphics{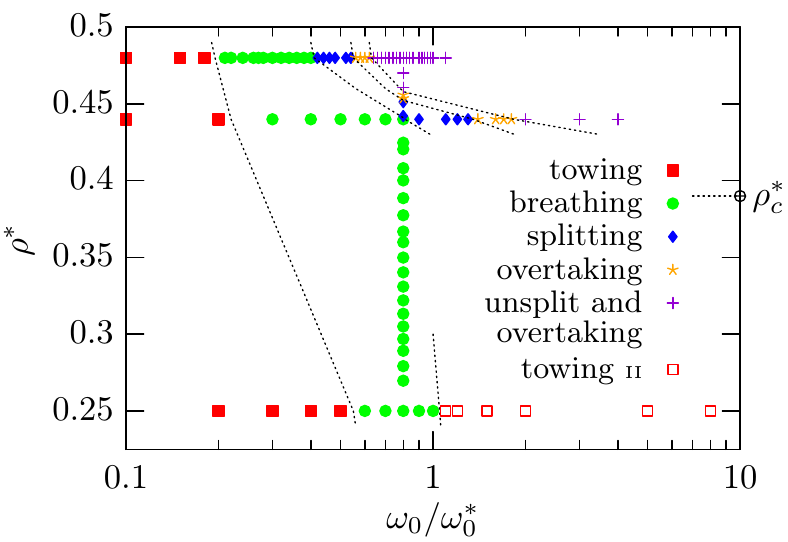}
\caption{Dynamical state diagram over driving frequency of the external potential $\omega_0$ in units of the critical frequency $\omega_0^*$ and reduced particle density $\rho^*$. Dashed lines serve as guide for the eye. %governed by
For descriptions of states see text.\label{fig:dynDiagram}}
\end{figure}

Our findings are summed up in the nonequilibrium state diagram depicted in Fig.~\ref{fig:dynDiagram}. The dashed lines separating different states are a guide for the eye.
The closer the system's parameters are to these boundaries the harder it is to classify its state unambiguously as it switches back and forth between neighboring states. 
We attribute this to the finite size of the system. 
The states given in Fig.~\ref{fig:dynDiagram} correspond to the predominant state for each point in $\rho^*$ and $\omega_0$.
\\
If we compare the dynamical state digram to the one calculated by dynamical fundamental measure density-functional theory by H\"artel et al.~\cite{Haertel_Towing:2010}, we find a very good qualitative agreement at 
high densities. There we observe the same states in identical order. 
However, for the low density regime we are able to discriminate between two distinct kinds of towing with a breathing transition regime between them where density-functional theory only observes towing.
This is partly due to the fact that it is impossible to distinguish between single particle rotation (towing) and the rotation of the director based on soliton-like dynamics (towing~\RNum{2}) 
in density-functional theory.\\
In a quantitative comparison, we find that the frequency band in which the different states emerge is considerably broader and lies at far lower frequencies than reported by H\"artel et al.~\cite{Haertel_Towing:2010}. 
Part of this discrepancy could be due to inertial effects as these have 
been reported to influence the boundaries between states~\cite{Jager_Pattern:2011}. Clearly, in our Langevin dynamics simulation inertia plays a non-negligible role, although we tried to minimize it 
(see Sec.~\ref{sec:numerical}), while in Ref.~\onlinecite{Haertel_Towing:2010} 
the spherocylinders are assumed to perform a completely overdamped Brownian motion.
On the other hand, H\"artel et al. observe even for the high density systems that the towing regime extends to more than five times the critical frequency $\omega_0^*$. This is in direct contrast to 
our findings 
that already below the critical frequency towing is not possible due to particle interactions. Even more so it is in disagreement with the single particle assessment (Sec.~\ref{subsec:single}) according to
which towing is prohibited for frequencies above the critical frequency. 
In dynamical density functional theory, the approximation is made that the
two-particle distribution function in a non-equilibrium system is equal to the equilibrium one
for the same density profile. In our system, particles collide more often with each other due to the time-dependent
external torques than they would in a system with a time-independent external field.
As a result, the positional dependence of the two-particle distribution function
should change compared to the equilibrium result at the same density profile and
the assumption made in DDFT cannot be expected to hold. Therefore, it is surprising that the DDFT does
correctly predicts which phases appear in this system.
%This suggest that while dynamical fundamental measure density-functional theory is able to predict the various dynamical states
%the positions of the boundaries between these can not be properly gained by that method.

%section Conclusion
\section{Conclusions} % and Future Work
\label{sec:conc}
We have simulated the behavior of a model colloidal liquid crystal consisting of hard spherocylinders under the influence of a periodic external potential that turns in a plane and acts on 
the orientation of the individual particles. We observe six different dynamical states that depend on the packing fraction of the particles and on the frequency of the external potential at 
fixed potential strength.
These states have been categorized by the characteristic behavior of the angular distribution of the particle orientations as towing, towing II, breathing, splitting, overtaking, and unsplit and overtaking.\\
At high packing fractions we find the same states as density-functional theory~\cite{Haertel_Towing:2010}. However, 
we observe a broad frequency range in which the dynamical states are found in contrast to the predictions of Ref.~\onlinecite{Haertel_Towing:2010}.
Moreover, we are able to distinguish between two distinct types of towing with an intermediate breathing regime at low packing fractions, which have both not been reported by density-functional theory. 
The quantitative comparison leads us to the conclusion
that with Langevin dynamics simulation we are able to provide a more reliable assessment of the state boundaries than given in previous studies.\\
We have analyzed the dynamical states we observe in detail and have given a thorough description from the single particle behavior point of view thereby explaining the underlying mechanisms that lead to 
the formation of those states.
The towing behavior observed at low driving frequencies is a state in which the rotation of each particle is synchronized with the external potential as observed for isolated 
particles~\cite{Shelton_Nonlinear:2005,Edwards_Synchronous:2006,Tierno_Overdamped:2009}. The dynamical states developing for higher frequencies originate from the interplay 
of the torque due to the external potential, friction with the fluid and particle-particle interactions. Here individual particles exhibit a periodic reversal of their turning direction, 
also seen for isolated particles at high frequencies~\cite{Shelton_Nonlinear:2005,Tierno_Overdamped:2009}. 
The fraction of particles that turn against the field which increases with both driving frequency and packing fraction (i.e. particle-particle interactions) determines the individual state. 
%(i.e. friction)

\begin{acknowledgments}
% Put your acknowledgments here.
The authors would like to thank Andreas H\"artel for useful discussions.
Financial support by the Deutsche Forschungsgemeinschaft (DFG) through
the Cluster of Excellence 'Engineering of Advanced Materials' in Erlangen,
and under grant Me1361/12 as part of the Research Unit
`Geometry and Physics of Spatial Random Systems',
is gratefully acknowledged.
\end{acknowledgments}

%\bibliography{bibfile}

\end{document}